\documentclass[prl, twocolumn,amsmath,amssymb]{revtex4}

\usepackage{hyperref}
\usepackage{graphicx}
\usepackage{dcolumn}
\usepackage{xcolor}
\usepackage{braket}
\usepackage{commath}
\usepackage{adjustbox}
\usepackage{comment} 
\usepackage{ulem}
\usepackage{tikz}
\usetikzlibrary{shapes}

\newcommand{\bs}[1]{{\boldsymbol{#1}}}
\newcommand{\bk}{\bs{k}}

\newcommand{\br}{\bs{r}}
\newcommand{\be}{\bs{\mathcal{\varepsilon}}}

\begin{document}

\title{Spin Hall effect of light in a random medium}

\author{Tamara Bardon-brun}
\affiliation{Laboratoire Kastler Brossel, Sorbonne Universit\'e, CNRS, ENS-PSL University, Coll\`ege de France; 4 Place Jussieu, 75005 Paris, France}

\author{Dominique Delande}
\affiliation{Laboratoire Kastler Brossel, Sorbonne Universit\'e, CNRS, ENS-PSL University, Coll\`ege de France; 4 Place Jussieu, 75005 Paris, France}

\author{Nicolas Cherroret}
\email{cherroret@lkb.upmc.fr}
\affiliation{Laboratoire Kastler Brossel, Sorbonne Universit\'e, CNRS, ENS-PSL University, Coll\`ege de France; 4 Place Jussieu, 75005 Paris, France}

\begin{abstract}
We show that optical beams propagating in transversally disordered materials exhibit a spin Hall effect and a spin-to-orbital conversion of  angular momentum as they deviate from paraxiality. 
We theoretically describe these phenomena on the basis of the microscopic statistical approach to light propagation in random media, and show that they can be detected via polarimetric measurements under realistic experimental conditions.
\end{abstract}


\maketitle

In random media, exploiting of the common wave-like nature of photons and electrons has led to the observation of several optical analogues of condensed-matter phenomena. 
Well known examples include fluctuations of photon conductance \cite{Scheffold98}, weak localization of light \cite{Albada85, Wolf85} or optical Anderson insulators \cite{Berry97, Schwartz07}. In Faraday active  materials, transverse diffusive currents resembling the Hall effect were also predicted \cite{Tiggelen95} and observed \cite{Rikken96}, and characterizations of photon localization under partially broken time-reversal symmetry were reported \cite{Bromberg16, Maret18}, in analogy with charged electrons in magnetic fields \cite{Beenakker97}. Recently, the propagation of paraxial light through disordered arrays of helical waveguides even allowed to realize a topological photonic Anderson insulator \cite{Stutzer18}. 

A currently open question is whether  spin-orbit interactions (SOI) of light, i.e.  the coupling between the spatial and polarization degrees of freedom of an optical wavefront, can be achieved in a random medium. A positive answer would be appealing, since in solids spin-orbit coupling is known to affect quantum transport,  giving rise for instance to weak anti-localization or driving random systems to other symmetry classes \cite{Evers08}. 
SOI could also be  used as a tool to design novel types of topological insulators \cite{Lu14, Ozawa18} in random environments.
As it turns out, optical SOI
naturally arise in inhomogeneous media. One of their manifestations is the optical spin Hall effect (SHE), which refers to helicity dependent sub-wavelength shifts of the  trajectory of circularly polarized beams, in analogy with their electronic counterparts \cite{Sinova15}. Originally identified 
for light refracted or reflected at interfaces (Imbert-Fedorov effect) \cite{Fedorov55, Imbert72} and later in gradient-index materials (optical Magnus effect) \cite{Dooghin92, Liberman92}, the SHE of light was recently described at a more general level on the basis of a geometric Berry phase 
\cite{Bliokh04, Onoda04}. On the experimental side, pioneering measurements at interfaces were carried out in optics \cite{Hosten08} and plasmonics \cite{Gorodetski12},
 and 
nowadays SOI of light have  become a promising tool for the generation of vortex beams or the optical control of nano-optical systems \cite{Cardano15, Bliokh15}. 
In this Letter, we demonstrate that the SHE of light is generically present in \textit{transversally} disordered media, a geometry recently exploited in the context of wave localization \cite{Boguslawski17, Schwartz07, Boguslawski13, Stutzer18}.
We find that the SHE emerges for beams deviating from the paraxial limit and exists even if the disorder is statistically homogeneous. While the SHE is naturally small, we show that it can be magnified and detected via polarimetric measurements under realistic experimental conditions.

\begin{figure}
\centering
\includegraphics[width=0.87\linewidth]{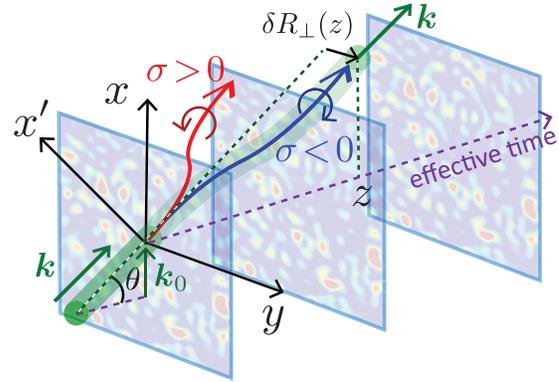}
\caption{
\label{SHE_scheme}
We consider the propagation of a collimated beam of wave vector $\bk$ through a medium spatially disordered in the $(x,y)$ plane   and homogeneous along the optical axis $z$.  As soon as the transverse wave vector $\bk_0$ (along the $x$-axis) is nonzero, the disorder-average centroid of the coherent mode is shifted  laterally along $y$ as $z$ increases (spin Hall effect). The shift is proportional to the beam helicity $\sigma$ and shows up even if the medium is  statistically homogeneous.
}
\end{figure}

In an inhomogeneous medium of permittivity distribution $\epsilon(\br)$, it was shown from semi-classical considerations that the mean coordinate $\textbf{R}$ and dimensionless wave vector $\textbf{P}=\bk c/\omega$ of an optical beam obey 
$\dot{\textbf{R}}=\textbf{P}/P-(\sigma c/\omega P^3)\textbf{P}\times\dot{\textbf{P}}$ and $\dot{\textbf{P}}=\bs{\nabla} {\sqrt{\epsilon(\boldsymbol{R})}}$, where $\sigma$ is the beam helicity, $\omega$ the optical frequency, $c$ the vacuum speed of light and the dot denotes derivation with respect to the optical path length \cite{Liberman92, Bliokh04, Onoda04}. The first 
equation of motion emphasizes the spin Hall effect of light, an helicity dependent, sub-wavelength spatial shift of the beam. Suppose now that $\epsilon(\br)$ describes a random medium. If the latter is statistically \textit{isotropic}, the beam momentum distribution and permittivity gradient are typically uncorrelated so that disorder averaging 
leads to $\dot{\textbf{R}}=\textbf{P}/P$
: no shift survives on average. To observe a finite optical SHE, a statistically \textit{anisotropic} disorder should be used. A simple configuration fulfilling this requirement 
is illustrated in Fig. \ref{SHE_scheme}: 
a monochromatic collimated beam, of wave vector $\bk$ lying in the $(x,z)$ plane, propagates in a material with disorder only in the transverse plane $\br_\perp=(x,y)$: $\epsilon(\br)=\epsilon(\br_\perp)$. This geometry has been much studied in the framework of the paraxial wave equation, in which the coordinate $z$ plays the role of an effective propagation time \cite{Raedt89}. By going beyond the paraxial description, we find  that beams 
carrying a finite helicity are \textit{laterally shifted} as soon as their transverse wave vector $\bk_0=k_0\textbf{e}_x$ is nonzero. This shift, which constitutes the optical SHE in a random medium, is visible in the so-called coherent mode, namely before the beam has been converted into a diffusive halo due to multiple scattering \cite{Sheng95}. 
Specifically, for an incoming beam of complex polarization vector $\be(z=0)=(\textbf{e}_{x'}+e^{i\phi}\textbf{e}_{y})/\sqrt{2}$, ($\textbf{e}_{x'}\equiv\textbf{e}_y\times\bk/k$),  we find a lateral shift (see Fig. \ref{SHE_scheme})
\begin{equation}
\delta R_\perp(z)=-\frac{\sigma}{k_0}\left[1-\frac{1}{\text{cosh}(z/2z_\text{SH})}\right]
\label{SHE_shift_circ}
\end{equation}
at small angle of incidence $\theta\simeq k_0/k\equiv \hat{k}_0$, with $\sigma=\sin\phi$ the beam helicity,  $\sigma=+1$ $(-1)$ for left(right)-handed circular polarization 
\cite{footnote2}.
The shift continuously increases as the beam propagates deeper in the random medium, until it saturates at $\sim 1/k_0$ beyond a characteristic time $z_\text{SH}\equiv z_s/\hat{k}_0^2$, where $z_s$ is the scattering mean free time \cite{Cherroret18}. 
The SHE vanishes for linearly polarized light, $\sigma=0$.
Note here the peculiarities of the transverse disorder scheme: the spin Hall shift evolves in time and is on the order of the transverse wavelength $1/k_0$, and not $1/k$ like in conventional shifts at interfaces \cite{Bliokh15}. 

To demonstrate Eq. (\ref{SHE_shift_circ}), a possible strategy consists in directly averaging the aforementioned semi-classical equations over disorder configurations. 
Because this approach seems hardly generalizable to higher orders of perturbation theory however, we have used a more general vector wave treatment based on the exact optical Dyson equation in random media \cite{Sheng95, Stephen86, vanTiggelen96, Busch05}.
Within this framework, we consider the evolution of the coherent mode in a transversally disordered material illuminated at $z=0$ by a collimated beam of electric field profile $\textbf{E}(\br_\perp,z=0)$. To describe this evolution, we define the normalized intensity, $\hat{I}(\br_\perp,z)\equiv I(\br_\perp,z)/I_\text{tot}(z)$, where $I(\br_\perp,z)\equiv |\langle\textbf{E}(\br_\perp,z)\rangle|^2$, $I_\text{tot}(z)\equiv\int d^2\br_\perp |\langle\textbf{E}(\br_\perp,z)\rangle|^2$ and the brackets refer to disorder averaging. The normalization is  here introduced so to work with a conservative quantity, as in a random medium the intensity of the coherent mode decays exponentially beyond the scattering mean free path \cite{Sheng95}, an effect that we will discuss later on.  
The components $E_j$ ($j=x,y,z$) of the complex electric field at coordinate $z$ obey the Helmholtz equation
\begin{equation}
\left[\Delta\delta_{ij}\!-\!\nabla_i\nabla_j\!+\!k^2\delta_{ij}(1\!+\!\delta\epsilon(\br_\perp)/\bar{\epsilon})
\right]E_j(\boldsymbol{r}_\perp,z)=0.
\label{Helmholtz_eq}
\end{equation}
Disorder is here encoded in random permittivity fluctuations
$\delta\epsilon(\br_\perp)=\epsilon(\br_\perp)-\overline{\epsilon}$ around a mean value $\overline{\epsilon}$. 
We choose them Gaussian distributed and correlated according to the general form 
$\overline{\delta\epsilon(\br_\perp)\delta\epsilon(\br_\perp')}/{\bar{\epsilon}^2}=B(\br_\perp-\br_\perp')$, where $B$ is an isotropic decaying function. The disorder average field, $\langle E_i(\br_\perp,z)\rangle=\int d^2\bk_\perp/(2\pi)^2\langle t_{ij}(\bk_\perp,z)\rangle E_{j}(\bk_\perp,z\!=\!0)e^{i\bk_\perp\!\cdot \br_\perp}$, is 
governed by 
the average transmission coefficient 
of the medium, 
$\langle t_{ij}(\bk_\perp,z)\rangle=2 i (k^2-\bk_\perp^2)^{1/2} \langle G_{ij}(\bk_\perp,z)\rangle$, where ${G}_{ij}$ is the Green's tensor of Eq. (\ref{Helmholtz_eq}) \cite{Feng94}. 
To find its disorder average, we have diagonalized the
Dyson equation for its Fourier transform, $\langle\textbf{G}(\bk_\perp,k_z)\rangle=[\textbf{G}^{(0)}(\bk_\perp,k_z)^{-1}-\boldsymbol{\Sigma}(\bk_\perp,k_z)]^{-1}$, where $G^{(0)}_{ij}(\bk_\perp,k_z)=(\delta_{ij}-\hat{k}_i\hat{k}_j)/(k^2-\bk_\perp^2-k_z^2+i0^+)$ and the self-energy tensor is evaluated at the level of the Born approximation: $\Sigma_{ij}(\bk_\perp,k_z)=\int d^2\bk_\perp'/(2\pi)^2 B(\bk_\perp-\bk_\perp')G^{(0)}_{ij}(\bk'_\perp,k_z)$ \cite{Cherroret18}. 
To make the calculation concrete, we model the incident light by a collimated Gaussian  beam $\textbf{E}(\br_\perp,z=0)=2/(\sqrt{\pi}w_0) \exp(-2 r_\perp^2/w_0^2+i\bk_0\cdot\br_\perp)\be(z=0)$ with unit polarization vector $\be(z=0)=(\textbf{e}_{x'}+e^{i\phi}\textbf{e}_{y})/\sqrt{2}$ and waist $w_0$ such that  $w_0 k_0\gg 1$.
By  using this initial state and the solution of the Dyson equation for the Green's tensor, we find, at leading order in $\hat{k}_0\ll 1$ (small $\theta)$:
\begin{equation}
\label{shifted_profile}
\hat{I}(\br_\perp,z)=\hat{I}(\br_\perp-\textbf{R}_\perp(z),0).
\end{equation}
\begin{figure}
\centering
\includegraphics[width=1\linewidth]{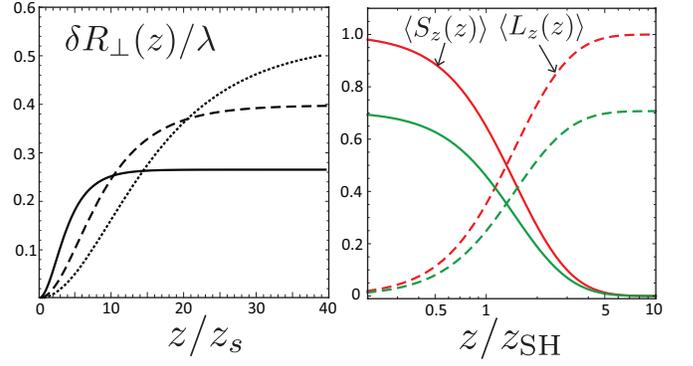}
\caption{
\label{StoO_conversion}
Left: spin Hall shift versus $z/z_s$ ($z_s$ mean free time) of a right-handed circularly polarized beam, for $\hat{k}_0=0.3$, $0.4$ and $0.6$ from top to bottom. Right: spin-to-orbital angular momentum conversion. As $z$ increases, the $z$-component of the spin angular momentum $\langle S_z(z)\rangle$ (solid curves) decreases and the beam acquires a finite orbital angular momentum $\langle L_z(z)\rangle$ (dashed curves), with the sum $\langle S_z(z)\rangle+\langle L_z(z)\rangle=1$ conserved. Upper and lower curves correspond to $\sigma=1$ and $0.7$, respectively.
}
\end{figure}
This result describes a shift of the centroid 
$\textbf{R}_\perp(z)\equiv \int d^2\br_\perp \br_\perp I(\br_\perp,z)/I_\text{tot}(z)$ 
as the beam evolves along the effective time axis $z$. The centroid shift is
\begin{equation}
\textbf{R}_\perp(z)=\hat{\boldsymbol{k}}_0z+\delta R_\perp(z)\bs{e}_y.
\label{SHE_shift_general}
\end{equation}
In Eq. (\ref{SHE_shift_general}), the first term on the right-hand side is the usual geometrical-optics contribution,
 while the second term is the 
 spin Hall effect of light, where $\delta R_\perp(z)$ is given by Eq. (\ref{SHE_shift_circ}) and the scattering mean free time follows from the angular average of the disorder power spectrum, $z_s^{-1}=k^3\langle B(k_0,\hat{\bk}_\perp\!\!-\!\!\hat{\bk}'_\perp)\rangle_{\hat{\bk}'_\perp}/4$ 
\cite{footnote1}. 
The left panel in Fig. \ref{StoO_conversion} shows $\delta R_\perp(z)$ versus $z$ in units of $\lambda=2\pi/k$, for three values of $\hat{k}_0$. Its asymptotic limit, $\delta R_\perp(z\gg z_\text{SH})=-\sigma/k_0$, increases with decreasing  ${k}_0$. Note that unlike the beam centroid, the mean momentum $\bk$ remains \textit{fixed} during propagation, as the coherent mode by definition describes the \textit{unscattered} part of the optical signal.

\begin{figure}
\centering
\includegraphics[width=0.8\linewidth]{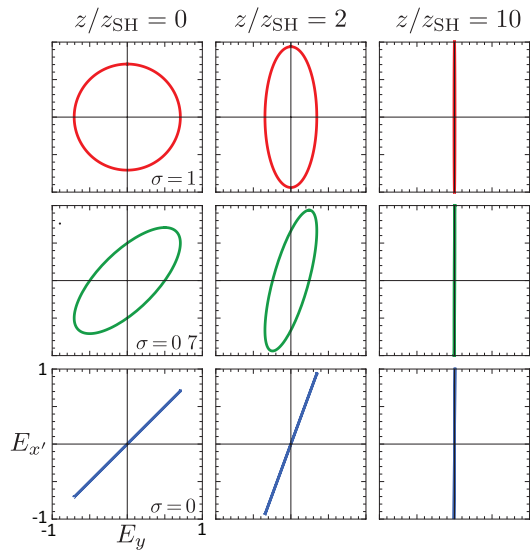}
\caption{
\label{Polarization_fig}
Real time evolution of the electric field vector $(E_{x'}, E_y)$ at increasing $z$, for $\sigma=1$ (top panels, circular polarization), $0.7$ (middle panels, elliptic polarization) and $0$ (lower panels, linear polarization). Whatever the initial polarization state, the beam always end up linearly polarized along $x'$ when $z\gg z_\text{SH}$.
}
\end{figure}

Often, optical SHE can alternatively be interpreted as a conversion of angular optical momentum: as the beam propagates in the inhomogeneous material, its spin angular momentum  is converted into an orbital angular momentum \cite{Bliokh15}. It turns out that, in a random medium, this picture holds for the coherent mode as well. To show this, we have  computed the angular momentum of the coherent mode, using the statistical approach described above. At small $\hat{k}_0$, the latter decomposes into an orbital contribution, $\langle\textbf{L}(z)\rangle\equiv-i\int d^2\br_\perp \langle E_i^*(\br_\perp,z)\rangle(\br\times\nabla)\langle E_i(\br_\perp,z)\rangle/I_\text{tot}(z)$, and a spin contribution, $\langle\textbf{S}(z)\rangle\equiv-i\int d^2\br_\perp \langle \textbf{E}^*(\br_\perp,z)\rangle\times\langle \textbf{E}(\br_\perp,z)\rangle/I_\text{tot}(z)$ \cite{Alonso12}. From the solution of the Dyson equation, we derive the transparent relation $\langle\textbf{L}(z)\rangle=\delta\textbf{R}_\perp(z)\times\bk$, which shows that the SHE can be regarded as the emergence of a finite orbital momentum. Of peculiar interest are the axial components $L_z(z)$ and $S_z(z)$, which explicitly read 
\begin{equation}
\langle L_z(z)\rangle\!=\!\sigma\!\left[1\!-\!\frac{1}{\text{cosh}(z/2z_\text{SH})}\right]\!,\ \langle S_z(z)\rangle\!=\!\sigma\!-\!\langle L_z(z)\rangle.
\end{equation}
 $\langle L_z(z)\rangle$ and $\langle S_z(z)\rangle$ are displayed in the right panel of Fig. \ref{StoO_conversion} as a function of $z$. As $z$ increases, the SOI mediated by the random medium convert $\langle S_z\rangle$ into $\langle L_z\rangle$ with no net angular momentum transferred to the random medium, which only acts as an intermediary. 
Note that unlike conversions previously reported in inhomogeneous anisotropic materials ($q$ plates) \cite{Marrucci06}, here  the spatial beam shape is preserved, see Eq. (\ref{shifted_profile}), so that the orbital angular momentum is ``external'', i.e. not associated with a vortex. In our system, the exact conservation of the $z$-component  of the total angular momentum stems from the global, statistical rotational symmetry around the $z$-axis.
 The spin-to-orbital conversion described here also indicates that the mean polarization of the incoming beam is not fixed, but evolves during propagation. From the Fourier component $\langle\textbf{E}(\bk_\perp\simeq\bk_0,z)\rangle\propto \be(z)$, we extract the explicit expression of the mean polarization vector $\be(z)$. For an initial beam with $\be(z=0)=(\textbf{e}_{x'}+e^{i\phi}\textbf{e}_{y})/\sqrt{2}$, we find:
\begin{equation}
\be(z)=\frac{\textbf{e}_{x'}+e^{i\phi}\exp(-z/2z_\text{SH})\textbf{e}_{y}}{\sqrt{1+\exp(-z/z_\text{SH})}}.
\label{polarization_evolution_eq}
\end{equation}
Trajectories of the electric field vector in the plane $(x',y)$ pertained  to Eq. (\ref{polarization_evolution_eq}) are represented in Fig. \ref{Polarization_fig} at increasing values of $z$ for $\sigma=1$, $0.7$ and $0$ (circular, elliptic and linear polarization, respectively). Due to the spin-to-orbital  angular momentum conversion,   the beam always end up linearly polarized along $x'$ beyond the spin Hall time $z_\text{SH}$, whatever the initial polarization state. Interestingly, the polarization vector of initially linearly polarized light rotates as well, although no SHE arises in this case. 

\begin{figure}
\centering
\includegraphics[width=0.95\linewidth]{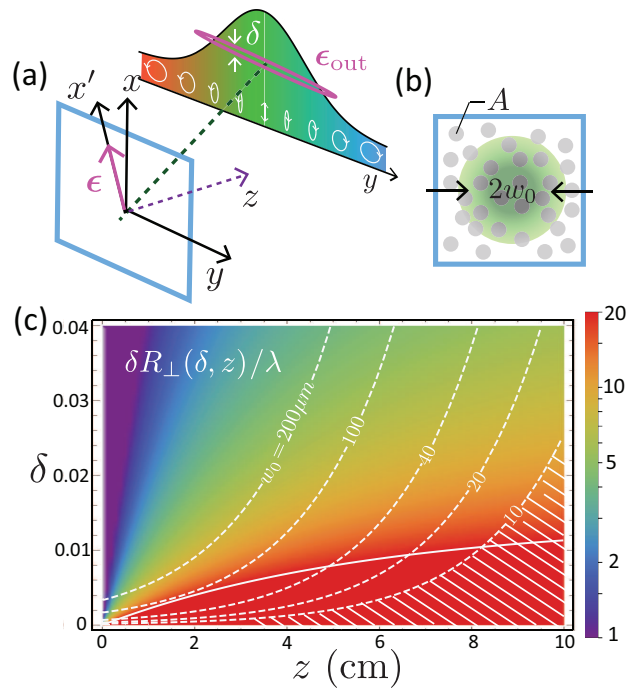}
\caption{
\label{Phase_diagram}
(a) When propagating in disorder, beams polarized along $\textbf{e}_{x'}$ acquire an inhomogeneous polarization structure with circularly polarized wings. By detecting light along $\be_\text{out}\propto(\textbf{e}_{y}+i\delta \textbf{e}_{x'})$ with $\delta\ll 1$, a magnified spin Hall shift can be observed. (b) To estimate the mean free time $z_s$, we consider a random array of guides of section $A$, surface density $\rho$ and relative refractive index $\delta n/n$. Taking $\lambda=532$ nm, $\delta n/n=7.10^{-4}$, $n=1.5$, $\hat{k}_0=0.57$, $A=20\mu$m$^2$ and $\rho=0.02\mu$m$^{-2}$, we find $z_s\simeq 1$ cm.
(c) Density plot of the spin Hall shift in wavelength units, Eq. (\ref{SHshift_polarimetric}), versus $z$ and $\delta$. 
Dashed white curves indicate the boundary where the constraint (\ref{criterion_SHE}) becomes an equality for various beam waists $w_0$. The SHE is only detectable in the region lying above these curves. The solid curve indicates the level set $\delta R_\perp=20\lambda$. 
}
\end{figure}

We finally present an experimental proposal for measuring the SHE of light in a random medium. As seen from Eq. (\ref{SHE_shift_circ}), in the transverse disorder scheme the maximum shift 
is $\sim 1/k_0$, which is much larger than $1/k$ at small angle of incidence. Unfortunately, this value is only reached at times $z>z_\text{SH}=z_s/\hat{k}_0^2$, typically longer than the mean free time $z_s$. In this regime, the signal  is exponentially attenuated because most photons have been scattered out of the initial mode and converted into a diffusive signal: $ I(\br_\perp,z)\propto \exp(-z/z_s)$ \cite{Akkermans07}. 
To circumvent this issue, a solution is to \textit{magnify} the SHE by means of a polarimetric measurement, in the spirit of previous works on optical shifts at interfaces \cite{Hosten08, Gorodetski12}. This strategy is based on a technique analogous to weak measurements in quantum mechanics \cite{Aharonov88, Duck89}
which, as we show now, can be applied to the coherent mode of a random medium as well. For this purpose, we suppose that the incoming beam is linearly polarized along $x'$, $\be(z=0)=\textbf{e}_{x'}$. 
In this particular configuration, the mean polarization $\be(z)$ remains fixed, but its spatial polarization \textit{distribution}, $\propto \langle \textbf{E}^*(\br_\perp,z)\rangle\times\langle \textbf{E}(\br_\perp,z)\rangle\propto \textbf{e}_{x'}\times\br_\perp\exp(-2r_\perp^2/w_0^2)$, is inhomogeneous: the core of the beam is linearly polarized while the wings $|\br_\perp|\sim w_0$ are circularly polarized with opposite helicities, as illustrated in Fig. \ref{Phase_diagram}(a). This implies that by detecting light at $z$ in the polarization channel $\be_\text{out}=(\textbf{e}_{y}+i\delta \textbf{e}_{x'})/\sqrt{1+\delta^2}$, with $\delta$ a small real number [Fig. \ref{Phase_diagram}(a)], a  spin Hall shift on the order of $w_0$ can be detected. Precisely, we find that the beam centroid, now defined as $\textbf{R}_\perp= \int d^2\br_\perp \br_\perp|\langle\be_\text{out}^*\cdot \textbf{E}(\br_\perp,z)\rangle|^2/\int d^2\br_\perp|\langle\be_\text{out}^*\cdot \textbf{E}(\br_\perp,z)\rangle|^2$, is given by:
\begin{equation}
\label{SHshift_polarimetric}
\delta R_\perp(\delta, z)\!=\!-\frac{\delta}{k_0}\frac{1\!-\!\exp(-z/2z_\text{SH})}{\delta^2\!+\!2[1\!-\!\exp(-z/2z_\text{SH})]^2/(w_0k_0)^2}.
\end{equation}
As compared to the case where no polarimetric measurement is performed, Eq. (\ref{SHE_shift_circ}), the SHE can now be enhanced by several orders of magnitude via the parameter $\delta$, with a maximum value 
$\delta R_\perp(z)\sim w_0$ for $z\gg z_\text{SH}$ and $\delta\simeq 1/w_0k_0$ \cite{Gorodetski12}.
To establish the practical conditions under which the SHE can be measured though, we must additionally account for the attenuation of the coherent mode. This attenuation is both due to the polarimetric measurement, which only selects a fraction $\delta^2$ of the intensity and, as discussed above, to multiple scattering. The latter depletes exponentially the coherent mode, which becomes weaker than the  diffusive signal emerging around after a few mean free times $z_s$.
This phenomenon, which constitutes the main limitation to a measurement of the SHE, imposes constraints on $z$ and $\delta$. To find them, we compare the intensity per unit surface of the coherent mode, $I\simeq \delta^2/(\pi w_0^2/2) \exp(-z/z_s)$ with $I_d$, the diffusive signal around. The latter was computed in \cite{Cherroret18} in the geometry of transverse disorder: $I_d\simeq [1-\exp(-z/z_p)]/(8\pi D z)$, where $z_p=8z_s/5\hat{k}_0^4$ and the diffusion coefficient $D=\hat{k}_0^2 z_s/2$ \cite{Cherroret18}. The constraint $I>I_d$ then reads:
\begin{equation}
\label{criterion_SHE}
\frac{w_0^2}{z z_s}[1-\exp(-z/z_p)]<8 \hat{k}_0^2\delta^2\exp(-z/z_s).
\end{equation}
From this criterion, it appears that the beam waist $w_0$ should be as small as possible.
For a realistic estimation,  we consider a medium consisting of a random array of uniformly distributed guides of surface density $\rho$, relative refractive index $\delta n/n$ 
and Gaussian profile of section $A$, $B(\bk_\perp)=\rho (A\delta n/n)^2\exp(-\bk_\perp^2A/4\pi)$, a type of disorder easy to imprint on glass \cite{Bellec12, Bellec17} [Fig. \ref{Phase_diagram}(b)].
With this model we find the mean free time from the Born approximation: $z_s^{-1}=\rho k^2(\delta n/n)^2A^{3/2}/4\hat{k}_0$. Using this expression, we show in Fig. \ref{Phase_diagram}(c) a density plot of the spin Hall shift versus  $(z,\delta)$, Eq. (\ref{SHshift_polarimetric}), obtained for $z_s\simeq 1$ cm. 
For a given waist $w_0$, the range of parameters where the inequality (\ref{criterion_SHE}) is satisfied lies above the dashed curve, as explicitly indicated by the shaded area for $w_0=10\mu$m. This analysis suggests that for $\delta\sim 10^{-2}$, a shift on the order of $20\lambda$ (level set indicated by the solid curve) could be detected for $w_0\sim10-40\mu$m. 

To conclude, we have demonstrated the spin Hall effect of light in transversally disordered media, starting from the general statistical treatment of wave propagation in random media. 
We have also proposed a practical experimental configuration where an amplified SHE can be detected via polarimetric measurements. Our study constitutes a first step toward a general description of SOI of light in random media, where they could be exploited to achieve novel regimes of wave transport or engineer gauge fields for photons.

NC thanks Cyriaque Genet and Matthieu Bellec for useful advice and comments.





\begin{thebibliography}{99}
\bibitem{footnote2}
This expression does not include the constant Imbert-Fedorov shift possibly arising at the interface between the outside and the random medium.
\bibitem{footnote1}
We here neglect the real part of the self energy, which shifts the average refractive index. We have checked than in usual transversally disordered media, in particular for the model in Fig. \ref{Phase_diagram}(b), this real part negligibly contributes to $\delta R_\perp(z)$.
\end{thebibliography}
\end{document}